\documentclass[twocolumn]{aastex701}
\usepackage[utf8]{inputenc}
\usepackage{amsmath}
\usepackage{ulem}
\usepackage{hyperref}
\usepackage{CJK}
\makeatother

\begin{document}
\begin{CJK*}{UTF8}{gbsn}

\title{H\,{\sc i} Gas and Star Formation in Major Galaxy Pairs from the FAST All-Sky H\,{\sc i} Survey (FASHI)}
\correspondingauthor{Taotao Fang} 
\email{fangt@xmu.edu.cn}
\author[orcid=0000-0002-3940-2950]{Shulan Yan (鄢淑澜)}
\email{yansl@stu.xmu.edu.cn}
\affiliation{Department of Astronomy, Xiamen University, Xiamen, Fujian 361005, People's Republic of China}
\author[orcid=0000-0003-3230-3981]{Qingzheng Yu (余清正)}
\email{qingzheng.yu@unifi.it}
\affiliation{Dipartimento di Fisica e Astronomia, Universit\`a degli Studi di Firenze, Via G. Sansone 1, 50019 Sesto Fiorentino, Firenze, Italy}
\author[orcid=0000-0002-2853-3808]{Taotao Fang (方陶陶)}
\email{fangt@xmu.edu.cn}
\affiliation{Department of Astronomy, Xiamen University, Xiamen, Fujian 361005, People's Republic of China}
\author[orcid=0000-0003-1761-5442]{Chuan He (何川)}
\email{hechuan@xmu.edu.cn}
\affiliation{Department of Astronomy, Xiamen University, Xiamen, Fujian 361005, People's Republic of China}
\author{Andrew Ma}
\email{andy.ma@student.isb.bj.edu.cn}
\affiliation{International School of Beijing, 10 An Hua Street, Shunyi District, Beijing 101318, People's Republic of China}
\author[orcid=0000-0003-4874-0369]{Junfeng Wang (王俊峰)}
\email{jfwang@xmu.edu.cn}
\affiliation{Department of Astronomy, Xiamen University, Xiamen, Fujian 361005, People's Republic of China}
\author[orcid=0000-0002-1588-6700]{C.Kevin Xu (徐聪)}
\email{congxu@nao.cas.cn}
\affiliation{National Astronomical Observatories, Chinese Academy of Sciences (NAOC), Beijing 100101, People's Republic of China}
\affiliation{Chinese Academy of Sciences South America Center for Astronomy, National Astronomical Observatories, CAS, Beijing 100101, People's Republic of China}
\author[orcid=0000-0001-6083-956X]{Ming Zhu (朱明)}
\email{mz@nao.cas.cn}
\affiliation{National Astronomical Observatories, Chinese Academy of Sciences (NAOC), Beijing 100101, People's Republic of China}
\affiliation{Guizhou Radio Astronomical Observatory, Guizhou University, Guiyang 550000, China}
\author[orcid=0000-0002-1189-2855]{Weishan Zhu (朱维善)}
\email{zhuwshan5@mail.sysu.edu.cn}
\affiliation{School of Physics and Astronomy, Sun Yat-Sen University, Zhuhai campus, No. 2, Daxue Road, Zhuhai, Guangdong 519082, People's Republic of China}

\begin{abstract}
   Atomic hydrogen (H\,{\sc i}) plays a fundamental role in fueling star formation in galaxies. However, the behavior of H\,{\sc i} gas in interacting systems, particularly galaxy pairs, remains elusive. In this work, we investigate the H\,{\sc i} content of major mergers by cross-matching the extragalactic H\,{\sc i} catalog from the FAST All-Sky H\,{\sc i} Survey (FASHI) with a previously established sample of isolated galaxy pairs. With the superior sensitivity of FAST, we have constructed the largest sample of major mergers with H\,{\sc i} detections, consisting of $440$ galaxy pairs: $364$ spiral-spiral (S+S) and $76$ spiral-elliptical (S+E) systems. We examine the H\,{\sc i} gas fraction ($f_{\mathrm{HI}}$), star formation rate (SFR) and H\,{\sc i} star formation efficiency ($\mathrm{SFE_{HI}}=\mathrm{SFR}/M_{\rm HI}$) for individual galaxies in pairs. The control sample is matched in both stellar mass and redshift. We find that paired galaxies, particularly those in pairs with small projected separations ($d_{\mathrm{p}}<50\ h^{-1}\mathrm{kpc}$), exhibit systematically lower (by $8.8\%$) H\,{\sc i} gas fractions compared to the control galaxies. The SFR is enhanced for galaxies in S+S pairs. $\mathrm{SFE_{HI}}$ is $\sim15\%$ higher for galaxies in S+S pairs than in the control galaxies, while spiral galaxies in S+E pairs show no significant difference in $\mathrm{SFE_{HI}}$ compared to the control sample. These findings suggest that the merging process triggers efficient H\,{\sc i} gas depletion and enhances star formation, especially in close S+S pairs. Notably, our sample includes $26$ red spirals in paired systems. These galaxies exhibit H\,{\sc i} deficiency and suppressed star formation activity compared to the isolated galaxies, indicating that interactions may affect quiescent spirals differently, potentially due to mechanisms similar to ellipticals.
\end{abstract}

\section{Introduction}
Atomic gas (H\,{\sc i}) is a fundamental component in galaxies, serving as the primary fuel for molecular gas formation in the interstellar medium (ISM) and contributing to the cool ionized gas content of the circumgalactic medium (CGM) \citep{Lehner11,Borthakur15,Wang20a}. In interacting systems, H\,{\sc i} provides a valuable diagnostic tool for probing gas dynamics during the merging process. Galaxy mergers are important for the formation of massive galaxies, as they drive significant changes in gas dynamics and star formation \citep{Lacey93,Mo10,Patton13,Ferreira25}. H\,{\sc i} observations of nearby galaxy pairs and post-mergers reveal tidal structures and bridges, indicating the impact of gravitational interactions on gas distribution \citep{Hibbard99,Wang01,Koribalski04}. These interactions can trigger bar instabilities, driving inflows of gas from large radii toward the galactic center and fueling central star formation or active galactic nuclei (AGN) activity \citep{Barnes96,Mihos96,Cox06,Torrey12,Sparre22,Xu24}. \\

Many studies have also investigated the evolution of H\,{\sc i} gas fraction during the merging process, but the results are still not clear. Some observational data of pre- and post-mergers indicate an enhancement of H\,{\sc i} gas \citep{Casasola04,Janowiecki17,Dutta18,Ellison18}. Conversely, other studies have reported no significant difference in H\,{\sc i} content or even a decrease in gas fraction after interactions \citep{Hibbard96,Georgakakis00,Ellison15,Zuo18}. The evolution of gas content also influences star formation in galaxy pairs. \citet{Scudder15} studied $17$ galaxy pairs and found that SFR is enhanced with increasing H\,{\sc i} gas fraction. According to \citet{Yu22}, H\,{\sc i} may be depleted and transformed into $\mathrm{H_{2}}$ particularly at the pericentric stage, which is important for star formation. The star formation efficiency in H\,{\sc i} ($\mathrm{SFE_{HI}}$) is higher in galaxy pairs, especially for galaxy pairs composed of two spiral galaxies \citep{Zuo18}. Numerical simulations by \citet{Moreno19,Moreno21} demonstrate that close encounters increase the cool gas content and elevate star formation activity. \citet{Sparre22} shows that rapid gas flows supply star formation and dilute the metallicity. In contrast, interactions between passive galaxy pairs are ineffective triggering star formation during interacting \citep{Brown23}. \\

Several large-scale H\,{\sc i} sky surveys, for example, HIPASS \citep{Barnes01,Wong06} and ALFALFA \citep{Giovanelli05,Haynes11} have significantly expanded the census of extragalactic atomic gas in the local universe, providing crucial datasets for studying H\,{\sc i} morphology, kinematics, and its role in fueling star formation \citep{Meyer04,Catinella18,Haynes18,Wang20a}. Using the ALFALFA dataset, \citet{Bok19} analyzed $348$ close, gas rich galaxy pairs and found that these galaxy pairs exhibit asymmetries in H\,{\sc i} profile. Furthermore, \citet{Bok20} studied $282$ galaxy pairs and reported that pairs have more stochastic gas fractions and greater H\,{\sc i} deficiency compared to isolated galaxies. More recently, \citet{Huang24} identified $278$ paired galaxies ($76$ in major mergers) detected by WALLABY and found monotonic enhancement of the SFR in high-mass pairs due to tidal effects. For low-mass pairs, the SFR is initially suppressed but enhanced when their H\,{\sc i} disks overlap. These H\,{\sc i} surveys highlight the role of atomic gas in probing the dynamical and star-forming processes triggered by galaxy interactions. \\

The Five-hundred-meter Aperture Spherical radio Telescope (FAST) is a powerful single dish telescope. The FAST all sky H\,{\sc i} survey (FASHI) project is expected to detect over $100,000$ extragalactic H\,{\sc i} sources covering the sky in $-14^{\circ}<\mathrm{decl.}<+66^{\circ}$ up to $z\sim0.35$ \citep{Zhang24}. The FASHI survey has a frequency range of $1.0$-$1.5$ GHz, achieving a detection sensitivity of $\sim0.76\ \mathrm{mJy\ beam^{-1}}$ at a velocity resolution of $6.4\ \mathrm{km\ s^{-1}}$ \citep{Jiang19,Jiang20}. This represents a significant improvement over HIPASS, ALFALFA and WALLABY \citep{Barnes01,Wong06,Giovanelli05,Haynes11,Haynes18,Koribalski20}. We cross-match the FASHI survey with one of the largest optically selected samples of galaxy pairs, compiled by \citet{Feng19}. Using SDSS, LAMOST and GAMA spectroscopy, their
work significantly increased both the number and completeness of known galaxy pairs at $z\leq0.25$ over a sky area of $\sim7000\ \mathrm{deg^2}$. By combining this extensive optical catalog with H\,{\sc i} data from FASHI, our study enables a more statistically robust investigation of atomic gas properties in merging galaxies, thereby advancing our understanding of the role of H\,{\sc i} during the interaction process. \\

This paper is organized as follows. In Section $2$, we describe the sample selection of galaxy pairs and isolated galaxies used as the control sample. Section $3$ presents our main results on the evolution of H\,{\sc i} gas and SFR during interactions. Section $4$ provides further discussion of these findings, and Section $5$  summarizes the main results. Throughout this paper, we adopt the standard $\Lambda$-CDM cosmology with $\Omega_{m}=0.3$, $\Omega_{\Lambda}=0.7$, $H_0=70\ \mathrm{km\ s^{-1}Mpc^{-1}}$ and $h=H_0/(100\ \mathrm{km\ s^{-1}Mpc^{-1}})$.\\

\section{Method}

\begin{deluxetable*}{ccccccccccccc}
\tablenum{1}
\setlength{\tabcolsep}{0.75mm}
\tablecaption{Paired galaxies and their H\,{\sc i} properties \label{tab:h1pair}}
\label{tab:sample_detail}
\tabletypesize{\footnotesize}
\tablewidth{0pt}
\tablehead{
\colhead{Galaxy} & \colhead{FASHI} & \colhead{$\mathrm{R.A.}_{\mathrm{HI}}$} & \colhead{$\mathrm{Decl.}_{\mathrm{HI}}$} & \colhead{$d_{\mathrm{p}}$} &
 \colhead{log($M_{*}$)} & \colhead{log(SFR)}  & \colhead{$z_{\mathrm{gal}}$} & \colhead{$W_{\mathrm{50}}$} & \colhead{$S_{\mathrm{bf}}$} & \colhead{log($M_{\mathrm{HI}}$)} & \colhead{$z_{\mathrm{HI}}$} & \colhead{Type}\\
\colhead{ID} & \colhead{ID} & \colhead{(deg.)} & \colhead{(deg.)} & \colhead{($h^{-1}\mathrm{kpc}$)} &
 \colhead{($M_{\odot}$)} & \colhead{($M_{\odot}/\mathrm{yr}$)} & \colhead{} & \colhead{($\mathrm{km\ s^{-1}}$)} & \colhead{($\mathrm{mJy\ km\ s^{-1}}$)} &
 \colhead{($M_{\odot}$)} & \colhead{} & \colhead{}
}
\colnumbers
\startdata
J132921-004647 & 20230004551 & 202.3394 & -0.7762 & 186.39 & 9.01 & -0.95 & 0.0164 & 175.37 & 2388.98 & 9.38 & 0.0164 & S+S \\
J032531-060744 & 20230000840 & 51.378 & -6.1353 & 160.06 & 10.57 & 0.69 & 0.0348 & 158.05 & 747.74 & 9.54 & 0.0347 & S+S \\
J122251-033121 & 20230002427 & 185.7376 & -3.5118 & 93.04 & 11.14 & 0.39 & 0.0712 & 44.65 & 1134.83 & 10.3 & 0.0712 & S+S \\
J110408+640447 & 20230041499 & 165.976 & 64.0417 & 115.41 & 9.81 & -1.24 & 0.0323 & 216.14 & 1338.07 & 9.74 & 0.0322 & S+S \\
J081526+453531 & 20230027469 & 123.8721 & 45.6356 & 107.65 & 10.87 & 0.35 & 0.0510 & 320.55 & 3317.12 & 10.48 & 0.0506 & S+E \\
J082759+493325 & 20230031342 & 127.0019 & 49.5646 & 156.23 & 9.99 & 0.08 & 0.0532 & 240.7 & 762.73 & 9.89 & 0.0533 & S+S \\
J080703+454035 & 20230027566 & 121.7399 & 45.702 & 79.18 & 10.83 & 0.17 & 0.0494 & 207.81 & 930.27 & 9.93 & 0.0496 & S+S \\
J090531+560126 & 20230037727 & 136.3836 & 56.0207 & 98.53 & 9.71 & -0.36 & 0.0254 & 143.33 & 3458.98 & 9.97 & 0.0254 & S+E \\
J101118+600045 & 20230040354 & 152.82 & 60.0185 & 184.89 & 8.95 & -0.87 & 0.0222 & 191.76 & 892.46 & 9.27 & 0.0222 & S+S \\
J094608+583149 & 20230039386 & 146.5073 & 58.5284 & 51.60 & 9.39 & -0.00 & 0.0299 & 315.92 & 3244.36 & 10.08 & 0.0299 & S+S \\
... & ... & ... & ... & ... & ... & ... & ... & ... & ... & ... & ... & ...\\
\enddata
\tablecomments{The columns show (1) the ID of matched galaxy in galaxy pairs; (2) H\,{\sc i} ID in FASHI catalog; (3) R.A. of H\,{\sc i} in degrees; (4) Decl. of H\,{\sc i} in degrees; (5) projected separation of galaxy pairs in $h^{-1}\mathrm{kpc}$; (6) Stellar mass of the galaxy; (7) SFR of the galaxy; (8) Redshift of the galaxy; (9) Velocity width of the H\,{\sc i} line profile in $\mathrm{km\ s^{-1}}$ from FASHI catalog, measured at $50\%$ level of every peak by busy-function fitting; (10) Integrated H\,{\sc i} line flux density in $\mathrm{mJy\ km\ s^{-1}}$ by busy-function fitting from FASHI catalog; (11) H\,{\sc i} mass; (12) redshift of H\,{\sc i} from FASHI catalog; (13) type of galaxy pair.}
\end{deluxetable*}
\subsection{Optical Galaxy Pair Sample}
\label{sec:opt_pair}

For our study of H\,{\sc i} gas in galaxy pairs, we adopt the optical galaxy pair catalog from \citet{Feng19}. The parent sample of \citet{Feng19} is from the New York
University Value-Added Galaxy Catalog (NYU-VAGC) of the SDSS DR$7$ \citep{Blanton05,Abazajian09}. They obtain redshifts from the SDSS DR$14$ \citep{Abolfathi18}, LAMOST DR$5$ \citep{Cui12,Zhao12,Luo15} and GAMA DR$2$ \citep{Liske15}. The criteria of isolated galaxy pairs of \citet{Feng19} are: (1) the projected separation for member galaxies $10\ h^{-1} \mathrm{kpc}\leq d_{\mathrm{p}}\leq 200\ h^{-1} \mathrm{kpc}$, (2) the line-of-sight velocity difference $\left|\Delta v\right|\leq 500\ \mathrm{km\ s^{-1}}$, (3) each pair member only has one neighbor satisfying the above criteria. They obtained $46,510$ galaxy pairs with $r$-band magnitudes in the range $14 \leq m_r \leq 17.77$ and redshifts of $z \leq 0.25$. The average spectroscopic completeness is $82\%$ in their sample. The stellar mass and SFR are taken from the public catalog of MPA-JHU DR$7$ \footnote{\url{https://wwwmpa.mpa-garching.mpg.de/SDSS/DR7/}} \citep{Abazajian09} and the $GALEX$-SDSS-$WISE$ Legacy Catalog \citep[GSWLC-X2;][]{Salim16,Salim18}. We first searched for the available data in the GSWLC-X2 catalog. For galaxies not included in GSWLC-X2, we adopted the measurements from the MPA-JHU DR$7$ catalog. \\

\subsection{H I Data of Galaxy Pairs }
\label{sec:h1_pair}
We use the extragalactic H\,{\sc i} data from the FASHI project \citep{Zhang24}. The first data release of the FASHI project covers the sky regions $0^{\mathrm{h}}\leq\mathrm{RA}\leq17.3^{\mathrm{h}}$, $22^{\mathrm{h}}\leq\mathrm{RA}\leq24^{\mathrm{h}}$ and $-6^{\circ}\leq\mathrm{DEC}\leq0^{\circ}$, $30^{\circ}\leq\mathrm{DEC}\leq66^{\circ}$. FASHI has detected $41,741$ extragalactic H\,{\sc i} sources. FASHI is able to detect faint sources, but these detections may be incomplete near the sensitivity limit. To ensure the detection of robust H\,{\sc i} signals and the reliability of our analysis, we selected the targets with integrated H\,{\sc i} line flux density $S_{\mathrm{bf}}>500\ \mathrm{mJy\ km\ s^{-1}}$ (including $\sim75\%$ of all FASHI targets). $95\%$ of these sources have SNR higher than $10$. \\
\begin{figure}[h!]
    \centering
    \includegraphics[width=0.36\textwidth]{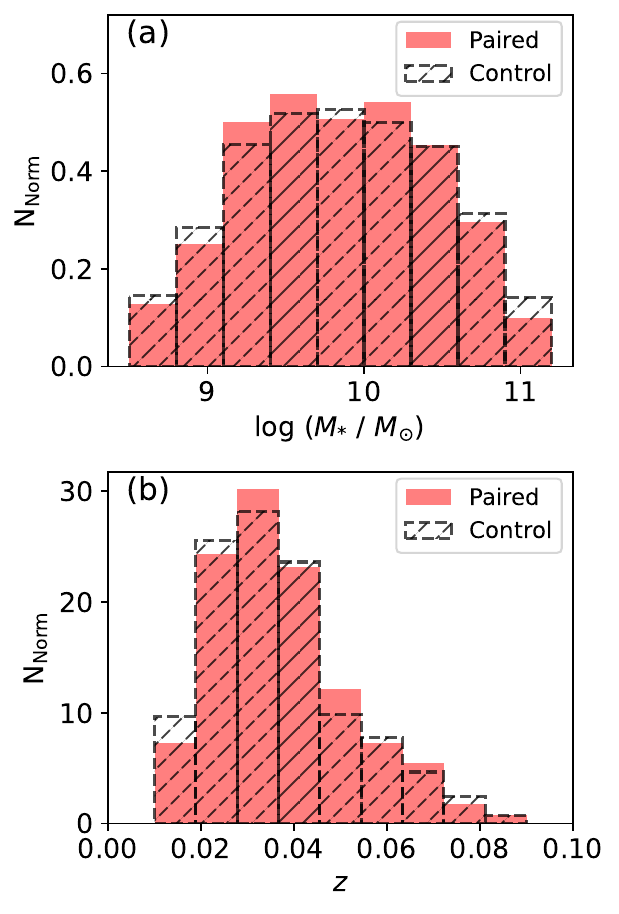}
    \caption{Distribution of the paired galaxies (red) and the control sample (black with slash), matched in mass (a) and redshift (b).}
    \label{fig:samples}
\end{figure}
\begin{figure*}[htbp]
    \centering
    \includegraphics[width=0.99\textwidth]{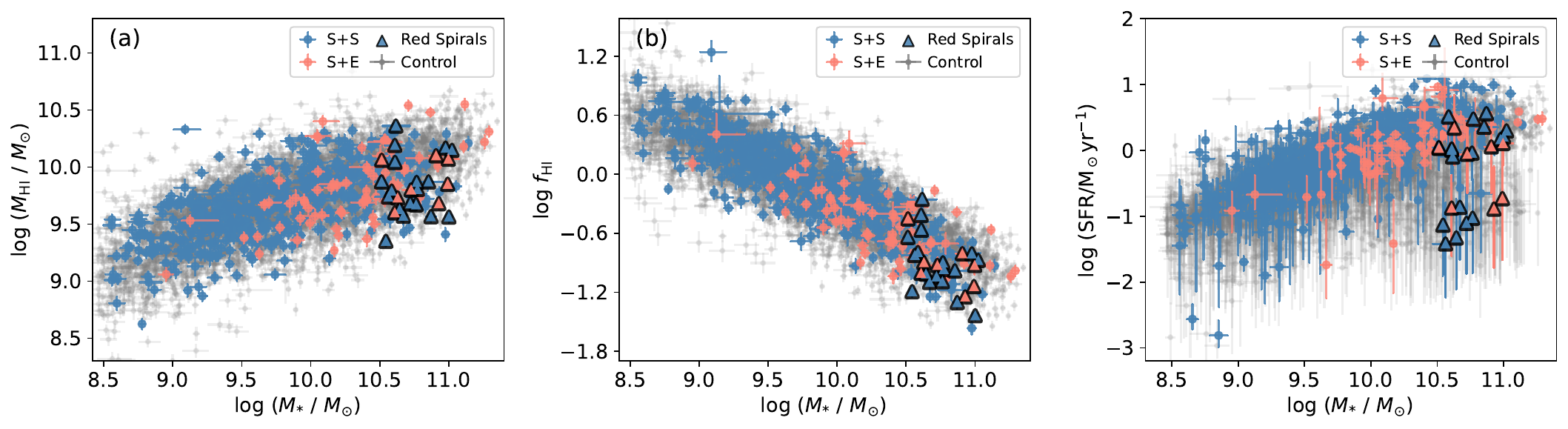}
    \caption{(a) H\,{\sc i} mass versus stellar mass. (b) H\,{\sc i} fraction versus stellar mass. (c) SFR versus stellar mass. Each point represents the S galaxy in one galaxy pair. The galaxies in S+S pairs are shown in blue dots, and those in S+E galaxies are shown in red dots, respectively. The triangles with black edges represent the red spirals. The control galaxies are shown using gray dots in the background. }
    \label{fig:HI_data}
\end{figure*}

To match FASHI H\,{\sc i} targets with paired galaxies within the beam size of FAST ($\sim2^{\prime}.9$), we searched for paired galaxies located within a $2^{\prime}.9$ radius of each H\,{\sc i} target's sky position. The velocity difference between the H\,{\sc i} target and galaxy pairs is $\Delta v\leq 1000\ \mathrm{km\ s^{-1}}$. We focus on galaxies with stellar mass $\mathrm{log}(M_*/M_{\odot})>8.5$ and major-merger pairs with mass ratio $\mu=M^1_{*}/M^2_{*}<3$, where $M^1_{*}$ and $M^2_{*}$ represent the stellar mass of the more and less massive members, respectively. A total of $440$ major-merger pairs have H\,{\sc i} detections, including $177$ unresolved pairs with angular separation $\leq2^{\prime}.9$ (the beam size of FAST). We classify the morphology type of galaxies based on S$\mathrm{\acute{e}}$rsic index $n_{S}$ \citep{Simard11}. We identify galaxies with S$\mathrm{\acute{e}}$rsic index $n_{S}<2.5$ as spirals \citep{Shen03}, and those with $n_{S}>2.5$ as ellipticals. For galaxies lacking a measurement of S$\mathrm{\acute{e}}$rsic index, we use visual classification data from the Galaxy Zoo project \citep{Lintott08,Lintott11}. We identify spiral galaxies with $P_{\mathrm{CS}}\ge0.6$ and others are ellipticals, where $P_{\mathrm{CS}}$ is the debiased probability of the spiral type. To avoid misidentifying young star-forming galaxies as ellpticals, we classify the galaxies with star formation rate $\mathrm{log(SFR/M_\odot\mathrm{yr^{-1}})}>0$ or specific star formation rate $\mathrm{log}(\mathrm{sSFR/yr^{-1}})>-11$ as spirals \citep{Popesso19,Elbaz11}. The final sample consists of $364$ spiral-spiral (S+S) types and $76$ spiral-elliptical (S+E) types. We consider individual galaxies in this work.   \\

The H\,{\sc i} mass is calculated via the formula \citep{Meyer04,Ellison18} : 
\begin{equation}
\frac{M_{\mathrm{HI}}}{M_\odot}=\frac{2.356\times10^{5}}{1+z}(\frac{D}{\mathrm{Mpc}})^{2}\frac{S_{\mathrm{bf}}}{\mathrm{Jy\ km\ s^{-1}}},
\label{eq:h1mass}
\end{equation}
where $D$ is the distance of source, $S_{\mathrm{bf}}$ is integrated H\,{\sc i} line flux density by busy-function fitting and $z$ is source redshift. We define H\,{\sc i} gas fraction using:
\begin{equation}
f_{\mathrm{HI}}=\frac{M_{\mathrm{HI}}}{M_{*}}.
\end{equation} 
The star formation efficiency of H\,{\sc i} gas ($\mathrm{SFE_{HI}}$) is defined as:
\begin{equation}
\mathrm{SFE_{HI}}=\frac{\mathrm{SFR}}{M_{\mathrm{HI}}}.
\label{eq:sfeh1}
\end{equation}
For unresolved pairs, we assume that H\,{\sc i} flux only comes from the spiral component for S+E pairs \citep{Zuo18}. For close S+S pairs, we assign the $S_{\mathrm{bf}}$ and H\,{\sc i} mass of each pair member by dividing them using $g$-band flux ratio. \citet{Denes14} reported a linear relation between H\,{\sc i} mass and optical/infrared magnitude, with the strongest correlation found for the $B$-band magnitude (Pearson coefficient $r=-0.81$). The effective wavelength of the $B$ band is $\lambda_{\mathrm{eff}}=4448\ \text{\AA}$ with a FWHM of $1008\ \text{\AA}$ \citep{Bessell90}, which is comparable to that of the SDSS $g$ band ($\lambda_{\mathrm{eff}}=4803\ \text{\AA}$, FWHM $=1409\ \text{\AA}$; \citet{Fukugita96}). We also tested the pairwise statistics for unresolved pairs in Appendix A and found no qualitative change to our results. To avoid uncertainties from faint sources, we only consider the S component in close S+S pairs with $S_{\mathrm{bf}}>500\ \mathrm{mJy\ km\ s^{-1}}$. In our final sample, we used the spiral galaxies with matched H\,{\sc i} detections in S+S pairs. For S+E pairs, only the S component is considered, as the E component is assumed to contribute negligibly to the H\,{\sc i} content. The final sample consists of $575$ spiral galaxies. The distribution of stellar mass and redshift for our paired galaxies is shown in red columns in Figure \ref{fig:samples}. Table \ref{tab:sample_detail} presents detailed information for ten example paired galaxies and their associated H\,{\sc i} properties. The full catalog is available online at \url{https://github.com/Yans59/HI_galaxy_pair}.\\

\subsection{Control Sample Matching}
To investigate the impact of galaxy interactions on atomic gas properties, we selected non-merger galaxies using the catalog from \citet{Feng19}. These galaxies are defined as galaxies without any bright neighbors ($m_r<17.77$) within $d_\mathrm{p}\leq200\ h^{-1}\ \mathrm{kpc}$ and $|\Delta v| \leq 500\ \mathrm{km\ s^{-1}}$ \citep{Feng20}. To obtain their H\,{\sc i} gas properties, we cross-matched non-merger galaxies with the FASHI catalog using the same method described in Section \ref{sec:h1_pair}. We also selected the galaxies with $S_{\mathrm{bf}}>500\ \mathrm{mJy\ km\ s^{-1}}$, resulting in a control pool of $5,546$ galaxies. To ensure a fair comparison and to account for the sensitivity of FASHI, each paired galaxy was matched with control galaxies based on stellar mass and redshift. Following \citet{Yu22} and \citet{Ellison18}, pairs and their controls were typically matched within the tolerance of $\left|\Delta \mathrm{log} (M_{\star}/M_{\odot}) \right|\leq 0.1$ dex and $\left|\Delta z \right|\leq 0.01$ dex. Furthermore, we ensured that the galaxy types of the control and paired galaxies were consistent. For each paired galaxy, we can match at least $7$ control galaxies. The median number of the control galaxies matched with each paired galaxy is 214. We also tested the results by matching the control galaxies while including SFR as an additional matching parameter. The evolution of the gas content remains unchanged when the SFR is included in the matching criteria. To better understand the evolution of SFR in galaxy pairs, we therefore adopt a matching scheme based only on the stellar mass, redshift, and the galaxy type. The distribution of the control sample is shown in black columns with a slash in Figure \ref{fig:samples}. \\

To make an accurate comparison of properties between paired and control galaxies, we calculate the offset of properties for each paired galaxy following \citet{Scudder12} and \citet{Yu22}:
\begin{flalign}
& \Delta f_{\mathrm{HI}} =\log f_{\mathrm{HI,pair}}-\log \mathrm{median}(f_{\mathrm{HI,controls}}),
\label{eq:dh1} \\
& \Delta\mathrm{SFR} =\log \mathrm{SFR_{pair}}-\log \mathrm{median}(\mathrm{SFR_{controls}}),
\label{eq:dsfr} \\
& \Delta\mathrm{SFE_{HI}} =\log \mathrm{SFE_{HI,pair}}-\log \mathrm{median}(\mathrm{SFE_{HI,controls}}),
\label{eq:dsfe}
\end{flalign}
where log median is the median value of H\,{\sc i} gas fraction, SFR, and H\,{\sc i} based SFE of matched control galaxies in the logarithm scale. Positive (negative) offset means enhancement (suppression) compared to the control sample. By matching the stellar masses and redshifts of paired and non-merger galaxies, we ensure that comparisons with the control sample are made under similar physical conditions. \\

\subsection{Derived Galaxy Properties and Classification}
Figure \ref{fig:HI_data}(a) and \ref{fig:HI_data}(b) present the H\,{\sc i} mass and H\,{\sc i} gas fraction versus galaxy stellar mass of paired galaxies, respectively. The control galaxies are shown by gray dots in each panel. Figure \ref{fig:HI_data}(a) and \ref{fig:HI_data}(b) reveal that the H\,{\sc i} mass increases and H\,{\sc i} gas fraction decreases with increasing stellar mass. The mean H\,{\sc i} mass shows no significant difference between the all paired and control galaxies, with $\mathrm{log}(M_{\mathrm{HI}}/M_{\odot})=9.70\pm0.01$ for both populations. All errors are $1\sigma$ standard error in the mean. We represent the galaxies in S+S pairs with blue points and those in S+E pairs with red points, respectively. Figure \ref{fig:HI_data}(a) and \ref{fig:HI_data}(b) show that all the paired galaxies follow the same trend as the control galaxies. \\

We represent log(SFR/$M_\odot \mathrm{yr^{-1}}$) versus galaxy stellar mass of the galaxy pairs in Figure \ref{fig:HI_data}(c). The mean SFR for all paired galaxies is $\log(\mathrm{SFR}/M_\odot \mathrm{yr^{-1}})=-0.14\pm0.02$. Isolated galaxies have $\log(\mathrm{SFR}/M_\odot \mathrm{yr^{-1}})=-0.24\pm0.01$. All paired galaxies exhibit the same trend as the control sample. However, they still show enhanced star formation rates compared to the control galaxies. \\

\begin{figure*}[ht!]
    \centering
    \includegraphics[width=0.99\textwidth]{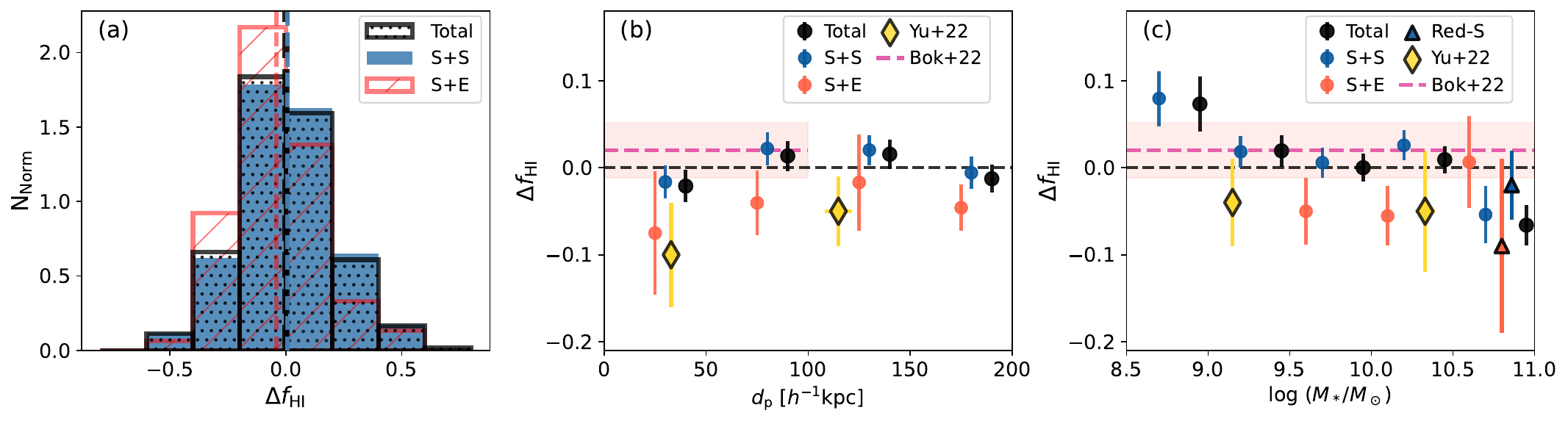}
    \caption{Panel (a) shows histograms of $\Delta f_{\mathrm{HI}}$ in paired galaxies. Black dotted columns represent $\Delta f_{\mathrm{HI}}$ for galaxies in all pairs, regardless of pair type. The galaxies in S+S and S+E galaxy pairs are indicated by blue and red shaded columns, respectively. The black, blue, and red dashed lines are the mean value of $\Delta f_{\mathrm{HI}}$ for the galaxies in all, S+S and S+E pairs, respectively. Panel (b) shows the evolution of $\Delta f_{\mathrm{HI}}$ at different projection separations ($d_{\mathrm{p}}$). Panel (c) represents $\Delta f_{\mathrm{HI}}$ against stellar mass. The galaxies in all pairs, regardless of pair type, are shown as black points with error bars in panels (b) and (c). The galaxies in S+S and S+E pairs of this work are indicated by blue and red points with error bar in (b) and (c), respectively. The black dashed line in each panel indicates the zero value. Red spirals in our sample are indicated by black-edged triangles in panel (c). The error bars represent the standard error of the mean.  Results from \citet{Yu22} are shown as yellow diamonds with black edges in panel (b) and (c). The pink dashed line with $1\sigma$ error region in panel (b) and (c) shows the results from \citet{Bok22}.}
    \label{fig:delta_fh1}
\end{figure*}

\section{Results}
\subsection{H I Gas Content in Galaxy Pairs}

Figure \ref{fig:delta_fh1}(a) shows the histogram of $\Delta f_{\mathrm{HI}}$. Total, the S galaxies in S+S and S+E pairs are marked with black, blue, and red, respectively. The full pair sample shows a mean $\Delta f_{\mathrm{HI}}=-0.001\pm0.01$ dex relative to control galaxies. The mean $\Delta f_{\mathrm{HI}}$ is $0.01\pm0.01$ for S galaxies in S+S pairs and $-0.04\pm0.02$ dex for those in S+E pairs, respectively. This corresponds to a $\sim9\%$ reduction in H\,{\sc i} gas fraction for S galaxies in S+E pairs, indicating a slight H\,{\sc i} deficiency in these systems. In contrast, the S galaxies in S+S pairs show no significant deviation from the control galaxies. \\

To investigate the evolution of H\,{\sc i} gas at different projection separations ($d_{\mathrm{p}}$), we divide our paired galaxies into $4$ sub-samples based on their projected separation. In Figure \ref{fig:delta_fh1}(b), we plot the $\Delta f_{\mathrm{HI}}$ of paired galaxies with $d_{\mathrm{p}}$. The data points for the S galaxies in total, S+S, and S+E pairs are represented by black, blue, and red points, respectively. The S components in S+S pairs dominate the overall trend and show no significant offset. They show a weak H\,{\sc i} deficiency in the closest bin, suggesting mild gas depletion at the smallest separations. The S galaxies in S+E pairs exhibit a clear trend of increasing H\,{\sc i} deficiency at smaller separations, with $\Delta f_{\mathrm{HI}}$ reaching $-0.08$ dex for $d_{\mathrm{p}}<50\ h^{-1}\mathrm{kpc}$. \\

The yellow diamonds with black edges show the results of paired galaxies from \citet{Yu22}, which comprises $\sim90\%$ H\,{\sc i} data from ALFALFA. They investigated $66$ galaxy pairs and found that $f_{\mathrm{HI}}$ in all paired galaxies is $\sim15\%$ deficient compared to control galaxies. At $d_{\mathrm{p}}<50\ h^{-1}\mathrm{kpc}$, the full paired galaxies in this work have modest deficiency in H\,{\sc i} fraction ($\Delta f_{\mathrm{HI}}=-0.04\pm0.02$, Figure \ref{fig:delta_fh1}(b)). However, the full paired galaxies with $d_{\mathrm{p}}>50\ h^{-1}\mathrm{kpc}$ have no significant H\,{\sc i} deficiency compared to \citet{Yu22}. One plausible factor is the difference in sensitivity between the surveys. Compared with ALFALFA, FASHI shows an integrated flux excess of $\sim10\%$ for bright sources (SNR$>40$), providing more accurate flux measurements for bright sources \citep{Zhang24}. H\,{\sc i} gas might be stripped outside the disk during the interaction. Therefore, part of the stronger suppression reported by \citet{Yu22} may arise from sensitivity limitations rather than intrinsic physical differences. FASHI is able to detect such extended structures, explaining the absence of measurable deficiency in galaxies of S+S pairs at $d_{\mathrm{p}}>50\ h^{-1}\mathrm{kpc}$. \\

Apart from H\,{\sc i} loss in major mergers, minor mergers can show H\,{\sc i} enhancement through gas accretion \citep{Janowiecki17,Lin25}. Although this work focuses exclusively on galaxy major mergers, our results also suggest possible differences in the H\,{\sc i} evolution between major and minor mergers. \citet{Casasola04} similarly found that H\,{\sc i} gas is enhanced in interacting systems, particularly in late-type galaxies. However, their sample did not account for the merger stage or distinguish between major and minor mergers. Overall, the evolution of H\,{\sc i} gas in galaxy pairs is complex. Different interaction stages, morphological types, and mass ratios can all lead to diverse outcomes. \\

H\,{\sc i} gas is more deficient in the S component of S+E pairs compared to the full sample. Early-type galaxies are typically gas-poor and have less-extended H\,{\sc i} disks, resulting in weaker hydrodynamic effects during interaction \citep{Moon19,Wang25}. Additionally, \citet{Moon19} suggests that the hot halo surrounding early-type galaxies may cut off cold gas accretion in their companions. This influence of the E component independent of redshift, local environment, stellar mass and mass ratio. For S component in our S+E pairs, the cold gas might be stripped and accretion of cold gas is suppressed when S component interacts with the halo gas of E component. As a result, S galaxies in S+E pairs have H\,{\sc i} deficiency even at $d_{\mathrm{p}}>150\ h^{-1}\mathrm{kpc}$. \\

We present $\Delta f_{\mathrm{HI}}$ with stellar mass in Figure \ref{fig:delta_fh1}(c). H\,{\sc i} deficiency is apparent in the high mass paired galaxies ($\log(M_*/M_{\odot})\ge9.5$).  We show the results of spiral paired galaxies in \citet{Bok22} using a pink dashed line in Figure \ref{fig:delta_fh1}(b) and (c). \citet{Bok22} studied $531$ paired galaxies with $d_{\mathrm{p}}<100\ h^{-1}\mathrm{kpc}$, including $377$ spirals. They found that deficiency of H\,{\sc i} content is more evident in elliptical paired galaxies, while spiral paired galaxies show no significant deficiency with $\Delta f_{\mathrm{HI}}=0.02\pm0.03$. Approximately $70\%$ of their paired galaxies have stellar mass in the range $9\le\mathrm{log}(M_*/M_{\odot})\le10.5$. Figure \ref{fig:delta_fh1}(c) indicates that $\Delta f_{\mathrm{HI}}$ of this work in the same mass range is consistent with the results of \citet{Bok22}. However, \citet{Bok22} did not consider the impact of projected separation or companion type, leading to uncertainties in the evolution of H\,{\sc i} gas during interactions. Low mass paired galaxies ($\log(M_*/M_{\odot})\le9$) in our sample show an enhancement in H\,{\sc i} gas fraction. It is probable that the FASHI survey detects more extended H\,{\sc i} disks than before. The enhanced H\,{\sc i} might originate from ongoing accretion but not migrate inward to fuel star formation. \\

We estimate the pair fraction as the ratio between the number of galaxy pairs and the number of isolated galaxies. For the full sample of \citet{Feng19}, the pair fraction is $\sim12.78\pm0.07\%$ at $z<0.1$. But it is only $\sim7.9\pm0.4\%$ in our H\,{\sc i}-selected sample. The difference suggests that a significant fraction of galaxy pairs might be HI-poor and likely fall the detection limit of FAST. Since our analysis primarily focuses on H\,{\sc i}-detected systems, it is inherently biased toward gas-rich galaxies. Therefore, the H\,{\sc i} content measured in our sample should be regarded as an upper limit on the typical H\,{\sc i} content of galaxy pairs, and the observed H\,{\sc i} deficiency relative to isolated galaxies likely represents a lower limit on the true effect of interactions. This result is consistent with the interpretation that gas depletion is a common outcome of galaxy interactions. \\

\begin{figure*}[ht!]
    \centering
    \includegraphics[width=0.99\textwidth]{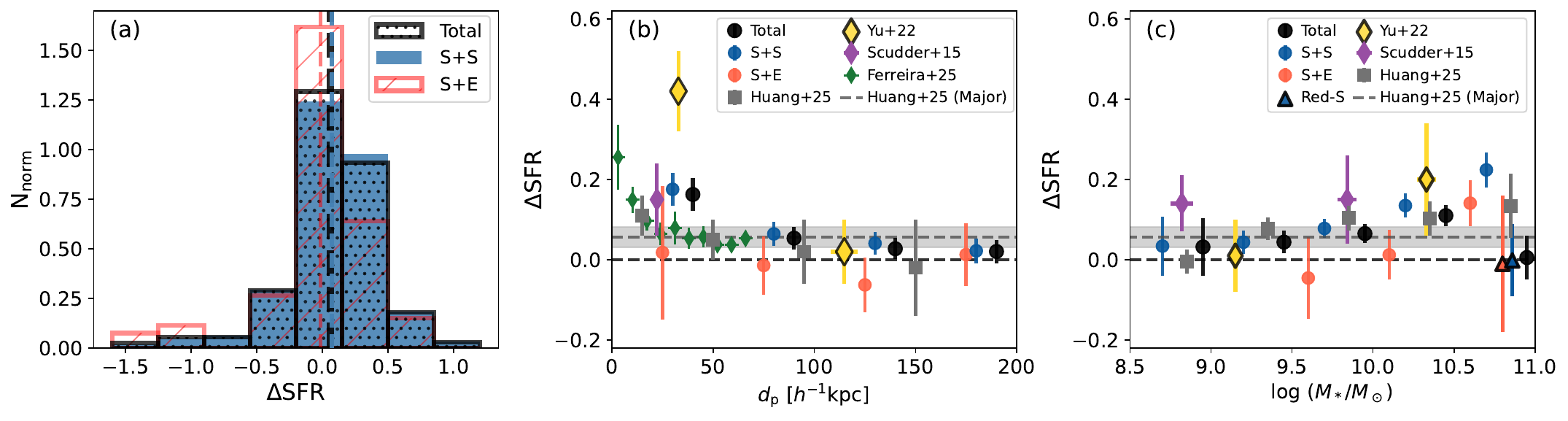}
    \caption{Panel (a) shows histograms of $\Delta$SFR in paired galaxies. Panel (b) shows the evolution of $\Delta$SFR in paired galaxies with different projection separation ($d_{\mathrm{p}}$). Panel (c) represents $\Delta$SFR against stellar mass. Results from \citet{Scudder15} and \citet{Huang24} are represented by purple diamonds and gray squares in panel (b) and (c), respectively. In panel (b) and (c), the gray dashed line with $1\sigma$ error region corresponds to $\Delta\mathrm{SFR}$ for major mergers from \citet{Huang24}. Green diamonds represent the results from \citet{Ferreira25}. Other color scheme and labels correspond to those used in Figure \ref{fig:delta_fh1}.}
    \label{fig:delta_sfr}
\end{figure*}

\subsection{Star Formation Enhancement in Major Mergers}

We present the distribution of $\Delta$SFR in Figure \ref{fig:delta_sfr}(a), where black, blue, and red columns represent the galaxies in the full sample, S+S pairs and S+E pairs, respectively. The full pair sample has the mean $\Delta\mathrm{SFR}=0.05\pm0.02$ dex. The mean $\Delta$SFR of galaxies in S+S and S+E pairs are $0.06\pm0.02$, $-0.03\pm0.04$ dex, respectively. The S galaxies in S+S pairs exhibit an enhancement of SFR ($\sim12\%$) when compared to isolated galaxies, while those in S+E pairs show no significant enhancement. \\

Figure \ref{fig:delta_sfr}(b) shows the trend of $\Delta$SFR with $d_{\mathrm{p}}$. S components in S+S pairs exhibit an increasing trend in $\Delta$SFR as $d_{\mathrm{p}}$ decreases. For close S+S pairs with $d_{\mathrm{p}}<50\ h^{-1}\mathrm{kpc}$, the SFR of these paired galaxies is enhanced obviously. $\Delta$SFR is $0.17\pm0.04$ dex in these paired galaxies, indicating that SFR is enhanced by $\sim48\%$ for galaxies in close S+S pairs.  \citet{Scudder15} investigated $17$ galaxy pairs with $d_{\mathrm{p}}<30\ h^{-1}\mathrm{kpc}$ and found an enhancement in star formation (purple diamonds in Figure \ref{fig:delta_sfr}(b) and (c)). The gray dashed line with a $1\sigma$ error region represents the mean $\Delta\mathrm{SFR}$ for $38$ major mergers of \citet{Huang24} ($\Delta\mathrm{SFR}=0.06\pm0.03$), which is consistent with the results of galaxies in S+S pairs in this work. \citet{Ferreira25} investigated the evolution of SFR from interacting systems to post-mergers using time-scale predictions, and we show their results for ongoing pairs as green diamonds. In their study, SFR shows a significant enhancement particularly at $d_{\mathrm{p}}<50\ h^{-1}\mathrm{kpc}$, up to $1.8\pm0.37$ times higher than in the control galaxies. \citet{Yu22} also observed an enhancement in star formation for pairs at $d_{\mathrm{p}}<50\ h^{-1}\mathrm{kpc}$ ($\Delta\mathrm{SFR}=0.42\pm0.10$). \citet{Yu22} distinguished merger stages using kinematic asymmetry, and close pairs in their sample are at the pericenter stage. But our sample may include a mix of all merger stages. To accurately distinguish the merger stages of our sample, future integral field unit data will be essential. \\

For paired galaxies with $d_{\mathrm{p}}>100\ h^{-1}\mathrm{kpc}$, both our sample and \citet{Huang24} show no significant enhancement in SFR. \citet{Patton13} reported a slight SFR enhancement even at such large separations (a factor $\sim1.3$), while \citet{Feng24} found that enhanced SFR at $d_{\mathrm{p}}>100\ h^{-1}\mathrm{kpc}$ occurs only in S+S pairs. However, the enhancement in \citet{Feng24} remains weak, with $\Delta\mathrm{SFR}<0.05$. Compared with previously optical-selected samples, the galaxy pairs in this work are relatively H\,{\sc i}-rich, suggesting that their atomic gas may not be efficiently converted into molecular gas or stars. In contrast, no significant enhancement of SFR is observed in our galaxies in S+E pairs, even in close pairs. These results indicate that the enhancement of SFR is more apparent in H\,{\sc i}-rich S+S pairs, especially at low projected separation. \\

In Figure \ref{fig:delta_sfr}(c), we plot the $\Delta$SFR against stellar mass. High-mass paired galaxies ($\log(M_*/M_{\odot})\ge9.5$) in this work and \citet{Yu22} exhibit an enhancement of SFR as shown in Figure \ref{fig:delta_sfr}(c). But pairs in \citet{Scudder15} show no variation of $\Delta\mathrm{SFR}$ with stellar mass. \citet{Scudder15} considered only $17$ close galaxy pairs, which may include bias due to sample selection. \citet{Huang24} also observed SFR enhancement in all massive paired galaxies (gray squares in Figure \ref{fig:delta_sfr}(b) and (c)). According to \citet{Huang24}, tidal interactions are stronger in high-mass galaxies, leading to gas inflow and enhanced star formation. The low H\,{\sc i} content may result from gas assumption for star formation and feedback process. While \citet{Huang24} focused on galaxy pairs regardless of mass ratio and morphology type, the H\,{\sc i} deficiency and SFR enhancement in this work is consistent with the tidal effects described by \citet{Huang24}. Further work considering minor mergers is necessary to study the impact of tidal effects in all types of galaxy pairs. \\

\section{Discussion}
\subsection{H I Gas, Star Formation, and Efficiency}

\label{sec:4_1}
\begin{figure}
    \centering
    \includegraphics[width=0.44\textwidth]{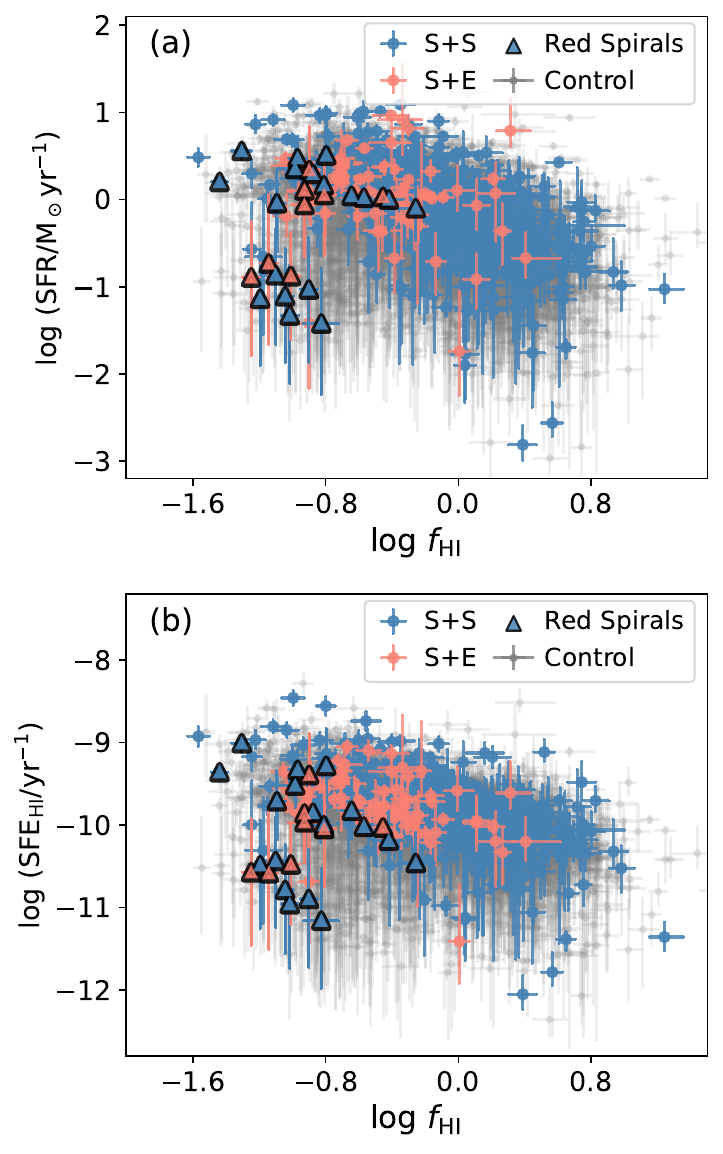}
    \caption{Plot of log(SFR) versus H\,{\sc i} fraction (a) and log($\mathrm{SFE_{HI}}$) versus H\,{\sc i} fraction (b). The color scheme and labels correspond to those used in Figure \ref{fig:HI_data}.}
    \label{fig:ssfr_h1}
\end{figure}

\begin{figure}[h!]
    \centering
    \includegraphics[width=0.44\textwidth]{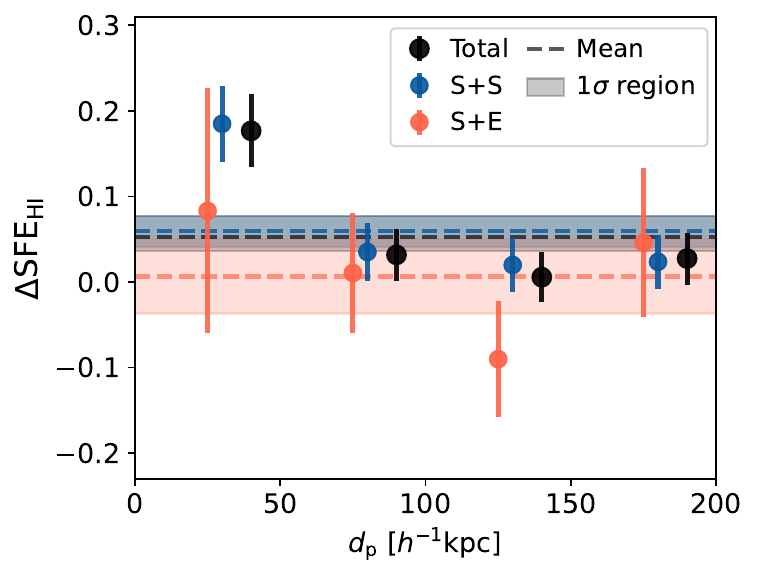}
    \caption{$\Delta\mathrm{SFE_{HI}}$ in different $d_{\mathrm{p}}$. We use black, blue and red points with error bars to indicate the galaxies in the total, S+S and S+E pairs, respectively. The dashed lines are the average $\Delta\mathrm{SFE_{HI}}$ for the galaxies in total, S+S and S+E pairs, respectively. The shaded region indicates $1\sigma$ error region.}
    \label{fig:delta_sfe}
\end{figure}

In this section, we discuss the evolution of H\,{\sc i} gas and star formation activity in galaxy pairs. Figure \ref{fig:ssfr_h1}(a) presents log(SFR) versus H\,{\sc i} gas fraction. Both galaxies in S+S and S+E pairs show decreasing SFR with increasing H\,{\sc i} gas fraction. But Figure \ref{fig:delta_sfr} shows that $\Delta\mathrm{SFR}$ of paired galaxies differs between S+S and S+E pairs. As mentioned by \citet{Zuo18}, gas content and the efficiency of converting gas into stars are possible key factors to explain the difference, rather than tidal interaction only. Previous CO observations have shown that molecular gas is more directly correlated with star formation \citep{Pan18,Violino18,Lisenfeld19,Yu24}. Although this work only focuses on atomic gas, it can still provide valuable insights into the gas content and star formation processes. We then calculate $\mathrm{SFE_{HI}}$ using Equation \ref{eq:sfeh1} and present it against H\,{\sc i} gas fraction in Figure \ref{fig:ssfr_h1}(b). $\mathrm{SFE_{HI}}$ in both pairs and the control sample decrease with increasing H\,{\sc i} gas fraction, suggesting that the efficiency of converting H\,{\sc i} gas into stars declines with increasing gas richness. As a result, the star formation rate is low for gas rich galaxies. \\

In Figure \ref{fig:delta_sfe}, we plot the $\Delta\mathrm{SFE_{HI}}$ against $d_{\mathrm{p}}$. The black, blue, and red dashed line indicates the mean value of $\Delta\mathrm{SFE_{HI}}$ in paired galaxies of total, S+S and S+E pairs, respectively. The mean $\Delta\mathrm{SFE_{HI}}$ is $0.05\pm0.02$ dex for all paired galaxies. The galaxies in S+S pairs have a mean $\Delta\mathrm{SFE_{HI}}=0.06\pm0.02$ dex, corresponding to a $15\%$ enhancement in $\mathrm{SFE_{HI}}$ compared to the control galaxies. For galaxies in close S+S pairs at $d_{\mathrm{p}}<50\ h^{-1}\mathrm{kpc}$, $\mathrm{SFE_{HI}}$ is enhanced by $\sim53\%$. But the S galaxies in S+E pairs have no significant enhancement of $\mathrm{SFE_{HI}}$ ($\Delta\mathrm{SFE_{HI}}=0.01\pm0.04$). This result suggests that the merging process may lead to H\,{\sc i} gas depletion to fuel star forming in close S+S pairs. \\

\begin{deluxetable}{ccc}
\setlength{\tabcolsep}{5mm}
\tablenum{2}
\tablecaption{Spearman's Rank Order Coefficient and the Significance of $\Delta\mathrm{SFR}$ versus $\Delta f_{\mathrm{HI}}$}
\label{tab:spear_sfr}
\tablehead{
\colhead{$d_{\mathrm{sep}} $($h^{-1}\mathrm{kpc}$)} & \colhead{S+S} & \colhead{S+E}}
\colnumbers
\startdata
0$\sim$50 & 0.02 (0.87) & 0.67 (0.05) \\
50$\sim$100 & 0.20 (0.01) & 0.17 (0.41) \\
100$\sim$150 & 0.10 (0.29) & 0.42 (0.08) \\
150$\sim$200 & 0.15 (0.10) & 0.07 (0.76) \\
\enddata
\tablecomments{Spearman's rank order coefficient $r_{s}$ with the $p$-value as significance in the parentheses among $\Delta\mathrm{SFR}$ and $\Delta f_{\mathrm{HI}}$. The columns are (1) subsample of galaxy pairs at different projected separation range. (2)-(4) Spearman's rank order coefficient and significance of the galaxies in S+S and S+E pairs.  }
\end{deluxetable}

To further investigate the relation between star formation and H\,{\sc i} gas, we performed a Spearman's rank order analysis among $\Delta\mathrm{SFR}$ and $\Delta f_{\mathrm{HI}}$ (Table \ref{tab:spear_sfr}). We use Spearman's rank order coefficient ($r_{s}$; by definition, $-1\le r_{s}\le1$) to quantify the goodness of the correlation. We consider a tight correlation with $|r_{s}|>0.6$ and a weak correlation with $0.3<|r_{s}|<0.6$. Conversely, we consider a correlation with $|r_{s}|<0.3$ as no correlation. The significance ($p$-value) is the possibility of the assumption that the null hypothesis is correct. As shown in Table \ref{tab:spear_sfr}, the correlation between $\Delta\mathrm{SFR}$ and $\Delta f_{\mathrm{HI}}$ is statistically insignificant for both S+S and S+E pairs, which is consistent with the results of \citet{Ellison18} and \citet{Yu22}. Figure \ref{fig:delta_fh1} and Figure \ref{fig:delta_sfr} demonstrate that H\,{\sc i} gas is depleted while SFR is enhanced in galaxy pairs. However, the relation between atomic gas and star formation is complex and influenced by multiple factors. One key factor is the molecular-to-atomic gas mass ratio, which plays a critical role in galaxy mergers. The depletion of H\,{\sc i} gas contributes to fueling the molecular gas. The enhancement of molecular gas has been observed in close galaxy pairs \citep{Pan18,Violino18,Lisenfeld19,Yu24}. Further observations targeting molecular gas are necessary to investigate the relationship between molecular and atomic gas in galaxy pairs. \\

\begin{figure}[h!]
    \centering
    \includegraphics[width=0.38\textwidth]{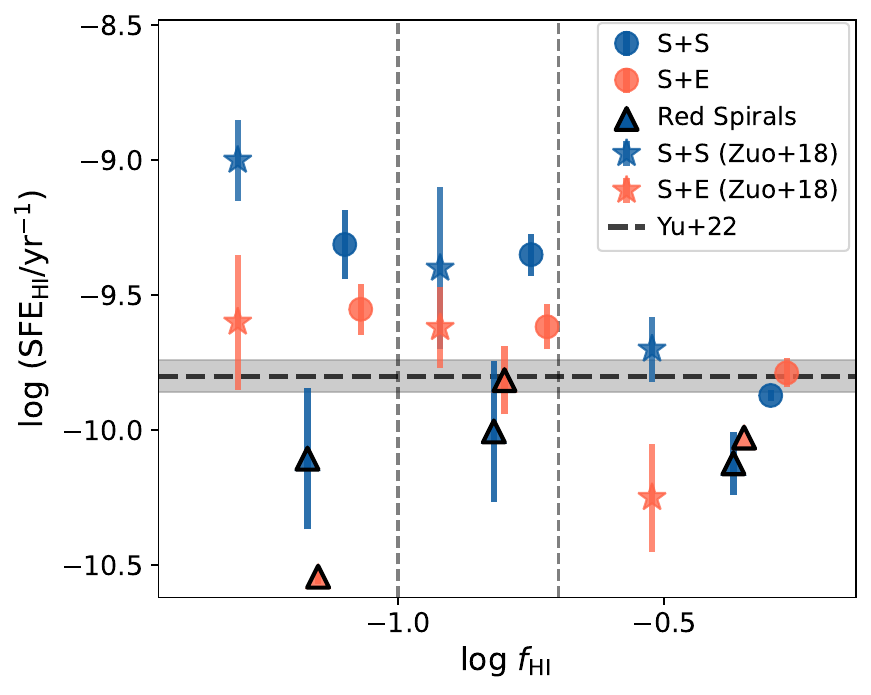}
    \caption{$\mathrm{SFE_{HI}}$ with H\,{\sc i} fraction. The blue and red circles represent the galaxies in S+S and S+E pairs without red spirals, respectively. The black-edge triangles show the results of galaxy pairs containing red-spirals. Stars indicate the results from \citet{Zuo18}. The results from \citet{Yu22} is shown by black dashed line with $1\sigma$ error region. The vertical gray dashed lines at $f_{\mathrm{HI}}=0.1$ and $0.2$ mark the boundaries between the 3 bins.}
    \label{fig:sfe_with_others}
\end{figure}

We also compared $\mathrm{SFE_{HI}}$ with previous studies. We calculate the average of $\mathrm{SFE_{HI}}$ in H\,{\sc i} fraction bins of $f_{\mathrm{HI}}<0.1$, $0.1<f_{\mathrm{HI}}<0.2$, $f_{\mathrm{HI}}>0.2$. Figure \ref{fig:sfe_with_others} shows $\mathrm{SFE_{HI}}$ in different $f_{\mathrm{HI}}$ bins. S+S pairs without red spirals are represented by blue circles, while S+E pairs without red spirals are represented by red circles. We show the results from \citet{Zuo18} using stars. The black dashed line with $1\sigma$ error region indicates the results of \citet{Yu22} ($\mathrm{log(SFE_{HI}/yr^{-1})}=-9.80\pm0.06$). The mean $\mathrm{log(SFE_{HI}/yr^{-1})}$ of all paired galaxies is $-9.84\pm0.02$ in this work, which is consistent with \citet{Yu22}. In general, $\mathrm{SFE_{HI}}$ decreases with increasing H\,{\sc i} gas fraction for S galaxies in S+S and S+E pairs, consistent with the findings of \citet{Zuo18}. The paired galaxies containing high $\mathrm{SFE_{HI}}$ may consume H\,{\sc i} gas more rapidly, leading to probable lower $f_\mathrm{HI}$ and higher $\mathrm{SFE_{HI}}$. \\

\subsection{Suppressed Star Formation in Red Spiral Galaxies}
\label{sec:4_2}

\begin{figure}[h!]
    \centering
    \includegraphics[width=0.38\textwidth]{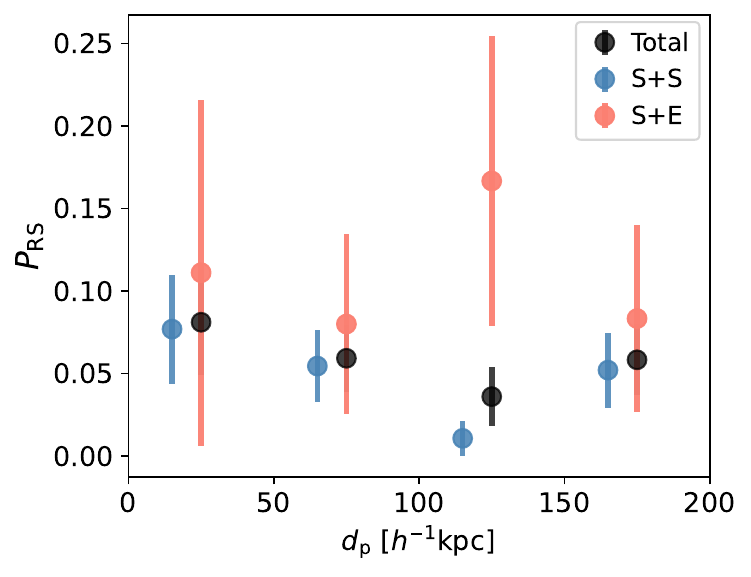}
    \caption{The proportion of galaxy pairs with red spirals at different $d_\mathrm{p}$. Red spirals in S+S (S+E) pairs are indicated using blue (red) dots, respectively. Black dots represent the proportion of all types. Error bars are binomial distribution uncertainty corresponding at $\pm1\sigma$ level.}
    \label{fig:pro_rs}
\end{figure}

In this work, we only consider the spiral galaxies in galaxy pairs. Recent studies have shown that a subset of spiral galaxies lies on the red sequence, commonly referred to as red spirals. Red spirals are typically characterized by low star formation activity despite their disk morphology \citep{Schawinski14,Guo16,Guo20}. Because of their quiescent nature, the red spirals may experience different evolution compared to those of normal spiral galaxies during interactions. To identify red spirals in our sample, we use the $u-r$ color mass relation from \citet{Guo16}: 
\begin{equation}
u-r>-0.24+0.25\ \mathrm{log}({M_*/M_{\odot}}).
\label{eq:red_s}
\end{equation}
We identified $26$ galaxy pairs containing red spirals, accounting for $\sim5.9\%$ of all pairs. Among these red spirals, $18$ are in S+S pairs and $8$ are in S+E pairs. We present red spirals using black-edge triangles in Figure \ref{fig:HI_data} and Figure \ref{fig:ssfr_h1}. The proportion of red spirals is $\sim4.9\%$ in S+S pairs and $\sim10.5\%$ in S+E pairs. Figure \ref{fig:pro_rs} presents the proportion of pairs containing red spirals in different $d_{\mathrm{p}}$ range. S+S and S+E pairs are shown using blue and red dots, respectively. The black dots indicate all types of pairs with red spirals. S+E pairs contain a higher fraction of red spirals ($\sim16.7\%$) compared to S+S pairs ($\sim1.1\%$) at $100<d_{\mathrm{p}}<150\ h^{-1}\mathrm{kpc}$, suggesting that red spirals in these pairs are more likely to have an early-type companion. \citet{Moon19} found that the suppression of SFR in S+E pairs can begin at high projected separations ($d_{\mathrm{p}}>150\ h^{-1}\mathrm{kpc}$), a scale comparable to the virial radius of Milky Way-sized galaxies. Red spirals in S+E pairs with $100<d_{\mathrm{p}}<150\ h^{-1}\mathrm{kpc}$ may be quenched due to early interactions with the E component.  \\

The average log(SFR) of red spirals is $-0.25\pm0.12\ M_\odot \mathrm{yr^{-1}}$. The mean H\,{\sc i} content of all red spirals is $\mathrm{log}\ f_{\mathrm{HI}}=-0.91\pm0.05$, corresponding to an H\,{\sc i} gas fraction of $\sim12.3\%$. \citet{Guo20,Wang22,Cui24} also found that H\,{\sc i}-rich red spirals can maintain H\,{\sc i} fractions exceeding $\sim10\%$, and up to $\sim60\%$. But the median H\,{\sc i} content in red spirals is lower than blue spirals. The observed H\,{\sc i} deficiency in our sample may reflect the intrinsically lower H\,{\sc i} content of red spirals relative to typical star-forming spirals. Compared to the control galaxies, the red spirals have a mean $\Delta f_{\mathrm{HI}} = -0.16\pm0.04$ dex and mean $\Delta \mathrm{SFR} = -0.50\pm0.12$ dex. During interactions, red spirals show a suppression in star formation. These galaxies also exhibit lower $\mathrm{SFE_{HI}}$ with a mean $\mathrm{log(SFE_{HI}/yr^{-1})}=-10.08\pm0.11$. Red spirals show different trends in Figure \ref{fig:sfe_with_others} compared to other samples. Disk galaxies with higher mass likely have higher angular momentum \citep{Mancera21}. Consequently, the high angular momentum of red spirals may reduce the inflow of gas into their central regions  \citep{Sancisi08,Mancera21b,Li24}. Figure \ref{fig:delta_fh1}(c) and Figure \ref{fig:delta_sfr}(c) show that red spirals are H\,{\sc i}-poor and star formation quiescent. As discussed in Section \ref{sec:4_2}, red spirals likely undergo a different evolutionary process than normal spirals. It is not surprising that red spirals show discrepancy in the trend observed in other samples. Red spirals may be inefficient at triggering star formation, even in the context of major mergers. 

\section{Summary}
We investigate H\,{\sc i} gas in nearby $440$ galaxy major mergers, including $364$ S+S and $76$ S+E pairs. The galaxy pairs were selected based on H\,{\sc i} detection by cross-matching pairs from \citet{Feng19} with the extragalactic H\,{\sc i} catalog from the FASHI project. To enable a precise comparison with isolated galaxies, we selected non-merger galaxies as a control sample. We focus on individual galaxies with H\,{\sc i} detection in pairs. For S+E pairs, we consider only the S component, since the E component contributes little to the H\,{\sc i} content. The final sample consists of $575$ spiral galaxies. We also identified $26$ paired red spirals following the selection method from \citet{Guo16}. The main conclusions of this work are as follows.  \\

1. For all paired galaxies, the H\,{\sc i} gas mass increases and H\,{\sc i} gas fraction decreases with increasing galaxy stellar mass. The mean $\Delta f_{\mathrm{HI}}$ values for galaxies in S+S and S+E pairs are $0.01\pm0.01$ and $-0.04\pm0.02$ dex, respectively. For the galaxies in close pairs with $d_{\mathrm{p}}<50\ h^{-1}\mathrm{kpc}$, those in S+S pairs have a mean $\Delta f_{\mathrm{HI}}=-0.03\pm0.02$, while those in S+E pairs have a mean $\Delta f_{\mathrm{HI}}=-0.09\pm0.06$. The deficiency of H\,{\sc i} content is more apparent in galaxies of S+E pairs. Halo interaction may play a crucial role in cold gas stripping in S+E pairs. \\

2. SFR shows an enhancement for galaxies in S+S pairs, with $\Delta\mathrm{SFR}=0.17\pm0.02$ dex. For close pairs with $d_{\mathrm{p}}<50\ h^{-1}\mathrm{kpc}$, the SFR for galaxies in S+S pairs is enhanced by $\sim48\%$. In contrast, the spiral galaxies in S+E pairs exhibit no enhancement of SFR and, in some cases, show suppression of SFR. We find no significant correlations of $\Delta f_{\mathrm{H\ I}}$ versus $\Delta\mathrm{SFR}$, consistent with the results from \citet{Yu22}. The evolution of $\Delta\mathrm{SFR}$ in our sample with $d_{\mathrm{p}}$ and stellar mass is consistent with the results of \citet{Huang24}, indicating that tidal effects in massive S+S pairs are important. \\

3. Using the criterion of \citet{Guo16}, we identified $26$ pairs containing red spirals in our sample, accounting for $\sim5.9\%$ of all pairs. The proportion of red spirals shows no significant relation with the projected separation. On average, red spirals contain $\sim12.3\%$ H\,{\sc i} gas content. The star formation is suppressed in red spirals with mean $\Delta \mathrm{SFR} = -0.50\pm0.12$ dex. \\

4. For all paired galaxies, the mean $\mathrm{log(SFE_{HI}/yr^{-1})}$ is $-9.84\pm0.02$. Our results suggest that $\mathrm{SFE_{HI}}$ is enhanced by $15\%$ for galaxies in S+S pairs compared to the control galaxies. The S galaxies in S+E pairs show no clear difference of $\mathrm{SFE_{HI}}$ with $\Delta\mathrm{SFE_{HI}}=0.01\pm0.04$. After excluding galaxy pairs with red spirals, we find a decrease in $\mathrm{SFE_{HI}}$ with increasing H\,{\sc i} gas fraction for galaxies in both S+S and S+E pairs. This trend is consistent with previous studies of \citet{Zuo18}.  \\

In conclusion, our study provides a larger sample of paired galaxies with H\,{\sc i} detection, enabling a detailed investigation of atomic gas evolution in interacting systems. We find that paired galaxies exhibit an evolution of atomic gas compared to the control galaxies. Additionally, star formation is affected during interactions. The evolution of H\,{\sc i} gas and SFR varies between different types of galaxy pairs. Galaxy pairs with different projected separations also have different properties. Despite these findings, the conversion between atomic gas and molecular gas in galaxy mergers remains unclear. Future observations focusing on molecular gas in galaxy pairs, based on our catalog, could provide deeper insights into the gas content in interacting systems. Future integral field unit data will allow us to distinguish the merger stages of galaxy pairs by analyzing their kinematic information. This will enable a more precise exploration of the physical properties of galaxy pairs at different stages of interaction. \\

We thank the helpful galaxy pair catalog derived by Shuai Feng. This work is supported by the National SKA Program of China No. 2025SKA0150103, National Natural Science Foundation of China under Nos. 12550002, 12133008, 12221003, 11890692. We acknowledge the science research grants from the China Manned Space Project with No. CMS-CSST-2021-A04 and No. CMS-CSST-2025-A10. QY was supported by the European Research Council (ERC) under grant agreement No. 101040751. JW acknowledges NSFC grants 12033004, 12333002, and China Manned Space Project CMS-CSST-2025-A07. This work uses the data from FASHI project. FASHI made use of the data from FAST (Five-hundred-meter Aperture Spherical radio Telescope).  FAST is a Chinese national mega-science facility, operated by National Astronomical Observatories, Chinese Academy of Sciences. \\

Funding for the Sloan Digital Sky Survey IV has been provided by the Alfred P. Sloan Foundation, the U.S. Department of Energy Office of Science, and the Participating Institutions. SDSS-IV acknowledges support and resources from the Center for High Performance Computing at the University of Utah. The SDSS website is \url{www.sdss.org}. \\

SDSS-IV is managed by the Astrophysical Research Consortium for the Participating Institutions of the SDSS Collaboration, including the Brazilian Participation Group, the Carnegie Institution for Science, Carnegie Mellon University, Harvard-Smithsonian Center for Astrophysics, the Chilean Participation Group, the French Participation Group, Instituto de Astrof\'isica de Canarias, The Johns Hopkins University, Kavli Institute for the Physics and Mathematics of the Universe (IPMU) / University of Tokyo, the Korean Participation Group, Lawrence Berkeley National Laboratory, Leibniz Institut f\"ur Astrophysik Potsdam (AIP),  Max-Planck-Institut f\"ur Astronomie (MPIA Heidelberg), Max-Planck-Institut f\"ur Astrophysik (MPA Garching), Max-Planck-Institut f\"ur Extraterrestrische Physik (MPE), National Astronomical Observatories of China, New Mexico State University, New York University, University of Notre Dame, Observat\'ario Nacional / MCTI, The Ohio State University, Pennsylvania State University, Shanghai Astronomical Observatory, United Kingdom Participation Group, Universidad Nacional Aut\'onoma de M\'exico, University of Arizona, University of Colorado Boulder, University of Oxford, University of Portsmouth, University of Utah, University of Virginia, University of Washington, University of Wisconsin, Vanderbilt University, and Yale University. \\

\software{Astropy \citep{astropy13},
SciPy \citep{scipy}, 
TOPCAT \citep{TOPCAT}}

\appendix

\section{Pairwise Statistic for Unresolved Pairs}

\begin{figure*}[ht]
    \centering
    \includegraphics[width=0.9\textwidth]{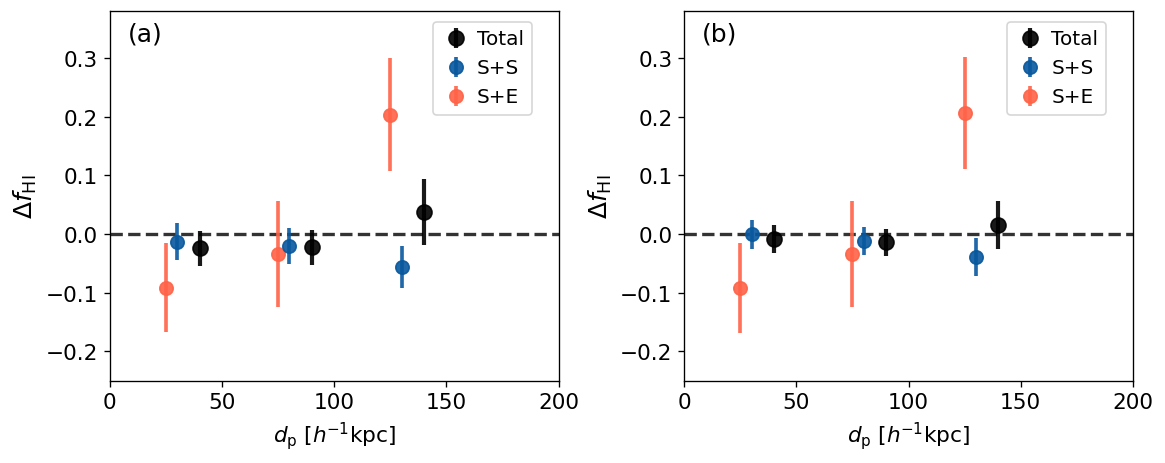}
    \caption{Panel (a) shows the evolution of $\Delta f_{\mathrm{HI}}$ in unresolved pairs without assignment at different projection separations ($d_{\mathrm{p}}$). Panel (b) represents the same evolution but with HI assignment. }
    \label{fig:delfh1_without_assign}
\end{figure*}

For unresolved galaxy pairs, we assigned the H\,{\sc i} flux based on the $g$-band flux ratio. We also performed a pairwise analysis for these unresolved systems, treating each pair as a single system. Correspondingly, we select two control galaxies and sum their H\,{\sc i} flux together. The results of $\Delta\ f_{\mathrm{HI}}$ are presented in Figure \ref{fig:delfh1_without_assign}(a). For comparison, our results under the H\,{\sc i} assignment method are presented in Figure \ref{fig:delfh1_without_assign}(b). The average $\Delta f_{\mathrm{HI}}$ is $-0.02\pm0.02$ for the full sample. When considering the pair type, the mean $\Delta f_{\mathrm{HI}}$ is $-0.02\pm0.02$ and $0.004\pm0.059$ for S+S and S+E pairs, respectively. We also show our previous results for unresolved pairs with HI assignment in Figure \ref{fig:delfh1_without_assign}(b). The mean $\Delta f_{\mathrm{HI}}$ for the galaxies in unresolved pairs is $-0.01\pm0.02$, $-0.01\pm0.02$, $0.004\pm0.059$ for the full sample, S+S pairs and S+E pairs, respectively. Figure \ref{fig:delfh1_without_assign}(a) and \ref{fig:delfh1_without_assign}(b) show very similar results for unresolved pairs, suggesting that our $\Delta f_{\mathrm{HI}}$ results are not driven by the assignment scheme.

\bibliography{references}{}
\bibliographystyle{aasjournal}

\end{CJK*}
\end{document}